\documentclass[aps,prb,twocolumn,shortbibliography,superscriptaddress]{revtex4-1}
\usepackage{epsfig}
\usepackage{epstopdf}
\usepackage{amsmath}
\usepackage{amsfonts}
\usepackage{amssymb}
\usepackage{hyperref}
\usepackage{bm}
\usepackage{makecell}
\usepackage{rotating}
\usepackage{hyperref}

\usepackage{graphicx}
\usepackage{dcolumn}
\usepackage{bm}
\usepackage{color}

\usepackage{tikz,xcolor,hyperref}

\definecolor{lime}{HTML}{A6CE39}
\DeclareRobustCommand{\orcidicon}{%
	\begin{tikzpicture}
	\draw[lime, fill=lime] (0,0)
	circle [radius=0.16]
	node[white] {{\fontfamily{qag}\selectfont \tiny ID}};
	\draw[white, fill=white] (-0.0625,0.095)
	circle [radius=0.007];
	\end{tikzpicture}
	\hspace{-2mm}
}

\foreach \x in {A, ..., Z}{%
	\expandafter\xdef\csname orcid\x\endcsname{\noexpand\href{https://orcid.org/\csname orcidauthor\x\endcsname}{\noexpand\orcidicon}}
}

\begin{document}

\title{Huge Dzyaloshinskii-Moriya interactions in Pt/Co/Re thin films}


\author{Amar Fakhredine\orcidB}
\email{amarf@ifpan.edu.pl}
\affiliation{Institute of Physics, Polish Academy of Sciences, Aleja Lotnik\'ow 32/46, PL-02668 Warsaw, Poland}
\affiliation{International Research Centre Magtop, Institute of Physics, Polish Academy of Sciences,
Aleja Lotnik\'ow 32/46, PL-02668 Warsaw, Poland}

\author{Andrzej Wawro}
\email{wawro@ifpan.edu.pl}
\affiliation{Institute of Physics, Polish Academy of Sciences, Aleja Lotnik\'ow 32/46, PL-02668 Warsaw, Poland}

\author{Carmine Autieri\orcidA}
\affiliation{International Research Centre Magtop, Institute of Physics, Polish Academy of Sciences,
Aleja Lotnik\'ow 32/46, PL-02668 Warsaw, Poland}

\date{\today}
\begin{abstract}
We investigate the magnetization and the Dzyaloshinskii-Moriya interactions (DMI) in Pt/Co/Re thin films in the case of perfect interfaces and upon the introduction of intermixing on both Co interfaces. Calculations were implemented on a series of systems with a varied number of cobalt atomic layers (ALs). Remarkably, the Re is able to introduce a DMI at the interface with cobalt and also, increase the DMI at the Pt/Co interface. We demonstrate that the chiral magnetic multilayer Pt/Co/Re with chiral spin structure can achieve a huge DMI value which is almost double of that attained in the prototype system W/Co/Pt. We study also the DMI as a function of the Re thickness finding the optimal thickness to maximize the DMI.
When we include a disorder that cancels a contribution from all first-neighbor Co atoms in the intermixed region, we found out that intermixing at the two interfaces affects the strength of the DMI solely when introduced at the Pt/Co interface where the DMI loses almost half of its value. On the contrary, the mixing at the Co/Re interface has very little or no effect as compared to the case with perfect interfaces. The value of the DMI would be somewhere between the perfect interface case and the totally intermixed case.
The realization of such a device should focus on the reduction of the Pt/Co intermixing to realize this huge DMI interaction.
\end{abstract}


\maketitle


A trilayer structure that exhibits the lack of inversion symmetry, with a ferromagnetic layer situated between two substantial heavy metals, potentially gives rise to intricate chiral magnetic configurations. At the core of this setup is the Dzyaloshinskii–Moriya interaction (DMI), characterized by an exchange energy term with an antisymmetric nature. This interaction takes on the role of directing the alignment of interacting spins in a non-collinear manner within systems lacking inversion symmetry and featuring robust spin–orbit coupling (SOC)\cite{moriya1960anisotropic, dzyaloshinsky1958thermodynamic, moriya1960new}. The DMI promotes for magnetic configurations with consistent rotational orientation, such as N\'eel-type domain walls and skyrmions \cite{heide2008dzyaloshinskii}. An example of this occurs at interfaces between cobalt and heavy materials like iridium or platinum, both of which possess significant SOC\cite{fert1990magnetic}. The skyrmions are topologically protected against continuous deformation towards spin structures with different topological invariants\cite{barker2016static, deng2022observation}
Skyrmions have technological applications in data storage, spintronics, logic devices, neuromorphic computing, quantum computing, magnetic sensors, and energy-efficient electronics\cite{10.1063/5.0042917}.
\begin{figure}[t!]
\centering
\includegraphics[width=9.9cm, angle=0]{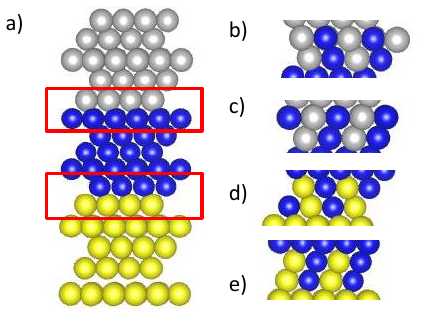}
\caption{(a) Layered structure of the Pt/Co/Re system with a perfect interface. The yellow, blue and grey balls represent the Pt, Co, and Re atoms, respectively. Crystal structures resembling the (b) Z-mixing and (c) C-mixing at the Pt/Co interface. Crystal structures resembling the (d) Z-mixing and (e) C-mixing at the Co/Re interface. The red rectangles are to indicate the area of two layers where the intermixing was introduced. After the intermixing, the number of layers composed entirely of cobalt atoms decreases by one in respect to the perfect interface.}\label{Structure}
\end{figure}

Several research papers have concentrated on investigating the DMI in Pt/Co bilayers, however, enhancing this chiral interaction could be achieved by sandwiching cobalt with another 5d material creating a chiral spin magnetic multilayer\cite{JIANG2017,10.1063/5.0021184}. One of the examples is the magnetic stack Pt/Co/W which was studied both experimentally and theoretically showing a DMI value that reaches higher than 2 mJ/m$^2$ \cite{jena2021interfacial}. On the other side, stacks like Al/Co/Pt and Ir/Co/Pt were reported in other works \cite{yang2018controlling} showing a weaker contribution that the bilayers Al/Co and Ir/Co can provide to the overall DMI. 
Our work shows that the use of another 5d material, that is, rhenium can induce a higher DMI value and be even more prospective.
For instance, it was reported that capping Co with a Re monolayer could give a clockwise contribution of around 1.5 meV nm, which is of the opposite sign to the contribution yielded by Pt/Co, which allows the system Pt/Co/Re to be a candidate for a huge DMI\cite{belabbes2016hund,jadaun2020microscopic}.
Recent experimental works on Pt/Co/Re have also confirmed this relatively large value and additive behavior of the DMI.\cite{nomura2022additive,Dhiman23,10228564} Even in bulk systems the rhenium largely increases the magnetic damping in FeCo alloy, demonstrating its effectiveness in inducing spin-orbit effects in Co-based system\cite{PhysRevB.99.174408}. \\
Many theoretical studies have validated the overall DMI of magnetic trilayers, which is usually overestimated with respect to experimental findings. The discrepancy observed can be attributed to the theoretical distinct emphasis on trilayers with sharp atomic layers (AL), while multilayers produced experimentally are anticipated to exhibit separate vertical textures due to intermixing effects at the interfaces.\cite{jia2020electric}. As a result, we also conducted calculations including interface disorder to seek an explanation for this type of disparity. \\
In this paper, we study the DMI in Pt/Co/Re thin films where we show the variation of this effect as a function of cobalt multilayers on a scale of up to 10 AL. The values of DMI achieved for our system exceed the prototype system for the chiral ferromagnetic multilayers W/Co/Pt\cite{RevModPhys.95.015003}. We find that the usage of Re actually affects the whole system and enhances the Pt/Co contribution even though it is not a direct neighbor to Pt, an effect which was also seen in the case of Rh\cite{jia2018first}. 
The origin of the system's huge DMI was found to be purely interfacial while the contributions to the DMI from the inner layers are oscillating and thus become non-effective. In addition, we study the effect of disorder at the two interfaces where we launched two different schemes of intermixed interfaces to indicate their effect on the magnetization and the DMI. We unveil a distinct robustness against the intermixing at the Pt/Co interface demonstrated by around 40\% reduction from the significantly high value achieved in the ideal case.\\


The technique of constraining the magnetic moments in a supercell was used to calculate the DMI of Pt/Co/Re multilayers through the density functional theory framework\cite{morshed2021tuning}. The Vienna {\it ab initio} simulation package (VASP) was used in all calculations using electron-core interactions described by the projector augmented wave method\cite{VASP}, and the exchange-correlation energy is calculated within the generalized gradient approximation of the Perdew-Burke-Ernzerhof (PBE) form\cite{perdew1996generalized}. The cutoff energies for the plane-wave basis set that were used to expand the Kohn-Sham orbitals were chosen to be 280 eV. The Monkhorst-Pack scheme was used for the $\Gamma$-centered 4$\times$12$\times$1 $k$-point sampling. 

A set of structural models of the Pt/Co/Re of the 4x1 unit cell stack were set up in a close-packed configuration. A primitive P1 space group unit cell was used to build the Pt/Co/Re multilayer stack supercells. For modeling the geometries of the interfaces, the experimental in-plane lattice constant of the Pt (111) surface which is 3.92 Å was used and hcp layer growth was assumed for Co and Re layers, while for Pt, fcc growth was considered. We varied the number of cobalt atomic layers ranging from 1 to 10 Als respective to each of our structures with a constant number of 5 AL for platinum and 5 AL for rhenium. The thin-film calculations were performed using slab geometry with a vacuum layer of 15 Å along the out-of-plane direction. This computational setup based on PBE has been used for the determination of magnetic properties in ferromagnetic multilayers\cite{autieri2016recipe,autieri2017systematic}. 

In order to extract the DMI vector, the calculations were carried out in three steps. First, structural relaxations were performed under a ferromagnetic state without SOC to determine the most stable interfacial geometries until the Hellman-Feynman forces converged to less than 0.015 eV/{\AA}. Next, the Kohn-Sham equations were solved without SOC, to find out the charge distribution of the system’s ground state. Finally, SOC was included and the self-consistent total energy of the system was determined as a function of the orientation of the magnetic moments which were controlled by using the constrained method implemented in VASP. The DMI was calculated using the method described by Yang et al.\cite{yang2015anatomy}. 

The DMI strength (d [meV]) parameter was calculated from the energy differences between clockwise and anti-clockwise energy configurations of the magnetic spirals in the cobalt layers for the perfect interface as shown in Fig. \ref{Structure}a). The DMI strength is further used to calculate the micromagnetic DMI which we use in our results. The micromagnetic DMI is calculated in units of mJ/m$^2$ and is a value that is inversely proportional to the number of ferromagnetic layers $N_F$ and the fcc lattice constant\cite{yang2015anatomy}. 
This method has been used for DMI calculations in bulk frustrated systems and insulating chiral-lattice magnets\cite{xiang2011predicting} and was adapted here in the case of interfaces\cite{yang2012strong}. In order to do so, we replaced two layers at the interface composed of 50\% Co and 50\% Pt or Re such that the number of total atoms is conserved relative to the case of perfect interfaces. We denote the types of mixing used for our calculations as C- and Z-type with respect to different atomic neighborhoods as shown in Fig. \ref{Structure}(b,c,d,e). The C- and Z-type have the same first-neighbor crystal structure but different second-neighbors, therefore, the results of the two calculations are expected to be close. We report both results to demonstrate the numerical accuracy of our DMI estimation. The Co and Pt(Re) are totally intermixed and there are no Co-Co first-neighbor in the intermixing region.

\begin{figure}[t!]
\centering
\includegraphics[width=\columnwidth,height=\columnwidth]{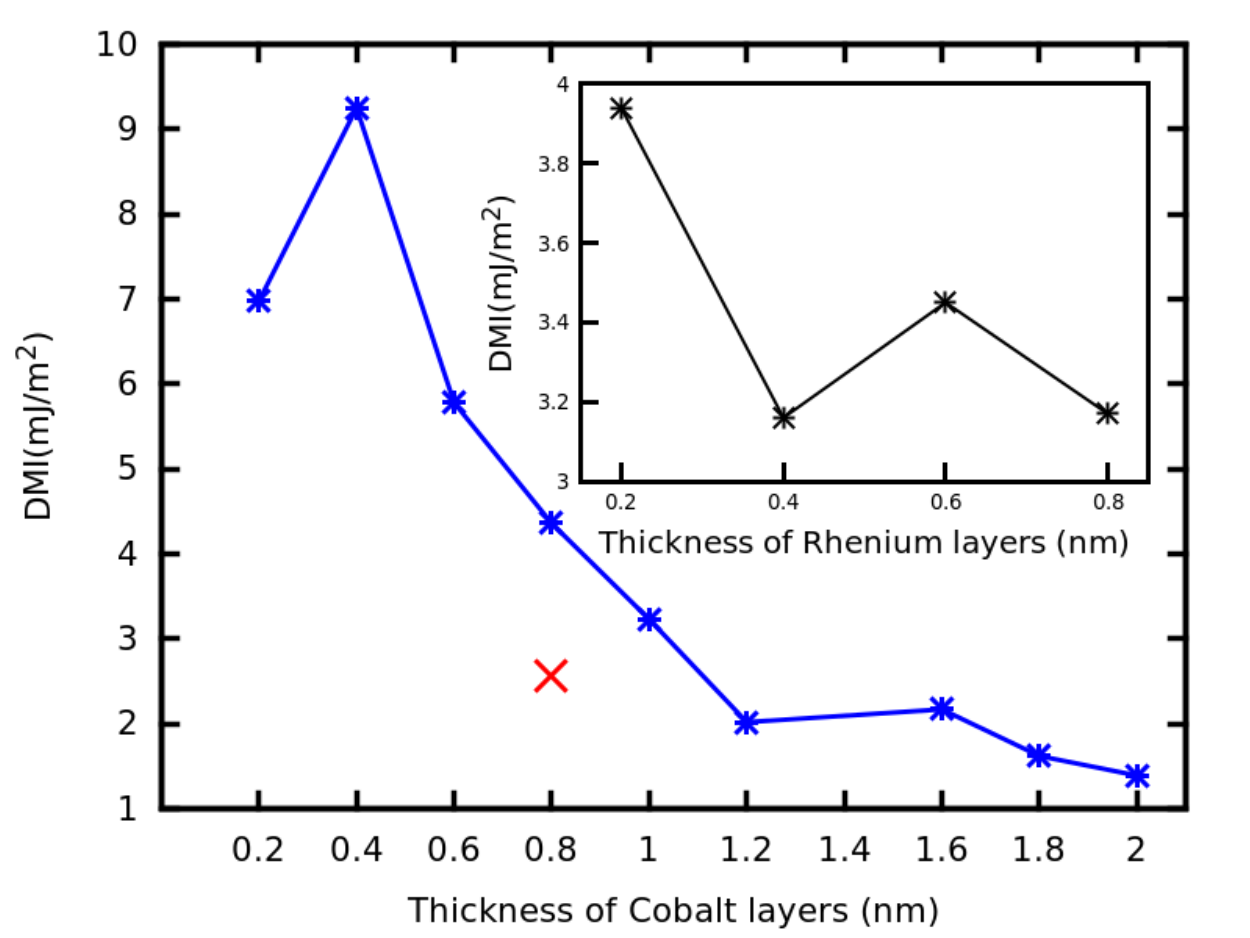}
\caption{Micromagnetic DMI results as a function of cobalt thickness (blue solid line) for Pt$_5$/Co(nm)/Re$_5$ where the subscript indicates the number of AL for a given element. Each cobalt AL has a thickness of 0.2 nm. For comparison, we add also the record-high value of W/Co/Pt from the literature\cite{jena2021interfacial,RevModPhys.95.015003} marked with a red cross. The inset of the figure shows the variation of the DMI as a function of rhenium layers for the thin film Pt$_5$/Co$_5$/Re(nm). The lines are a guide for the eyes.}\label{DMI}
\end{figure}

We divide our results into two sections, where we describe the system with perfect interfaces in the first part, while in the second part, we consider the system by adding mixed interfaces with variable selected thicknesses of cobalt ALs.


We calculate the micromagnetic DMI for the Pt/Co/Re thin film with perfect interfaces represented in Fig. \ref{Structure}a). 
The calculations of the total magnetic DMI contribution followed a declining trend as we increased the thickness of the cobalt layer as shown in Fig. \ref{DMI} reaching the maximum value of 9.25 mJ/m$^2$ at 2 AL and dropping to a value of 1.39 mJ/m$^2$ at a thickness of 10 ALs. The curve roughly follows the trend of $\frac{1}{N_F}$ for the micromagnetic DMI\cite{yang2015anatomy} with some oscillations due to the metallic nature of the systems.
The case of one cobalt layer is a peculiar point since the only Co-layer shares the same interface with rhenium on one side and platinum on the other side, producing a value of 6.98 mJ/m$^2$.
In the inset, we report the evolution of the DMI as a function of the Re thickness for the Pt$_5$/Co$_5$/Re(nm) thin films. 
In contrast to cobalt, changes in the arrangement of rhenium layers have a relatively smaller impact on the strength of the DMI. The DMI shows a gradual variation from 3 to 4 mJ/$m^2$ as the rhenium thickness increases from 1 to 4 ALs, which is close to the value calculated for the system Pt$_5$/Co$_5$/Re$_5$ that is 3.23 mJ/m$^2$.

\begin{figure}[t!]
\centering
\includegraphics[width=0.5\columnwidth,height=5cm]{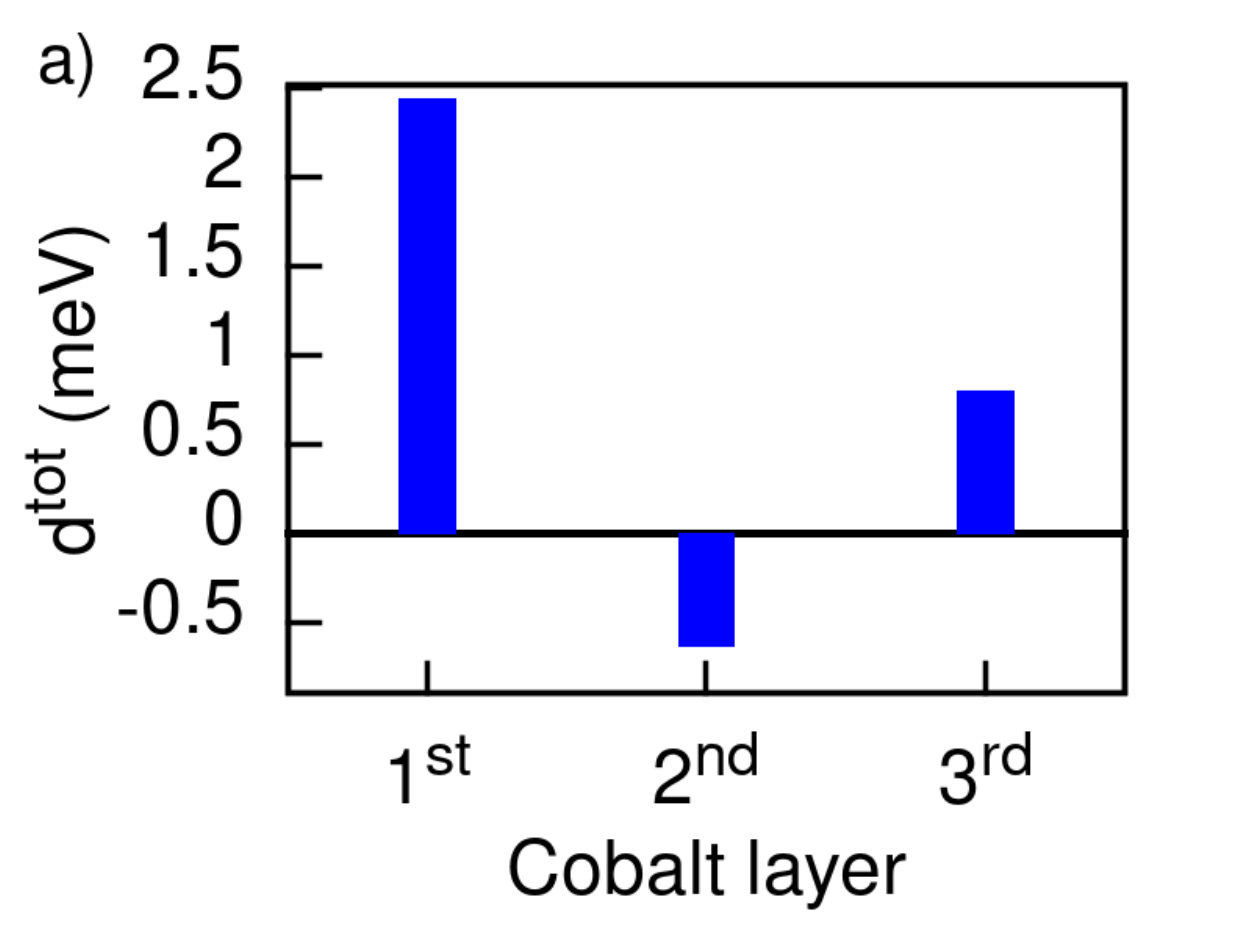}\includegraphics[width=0.5\columnwidth,height=5cm]{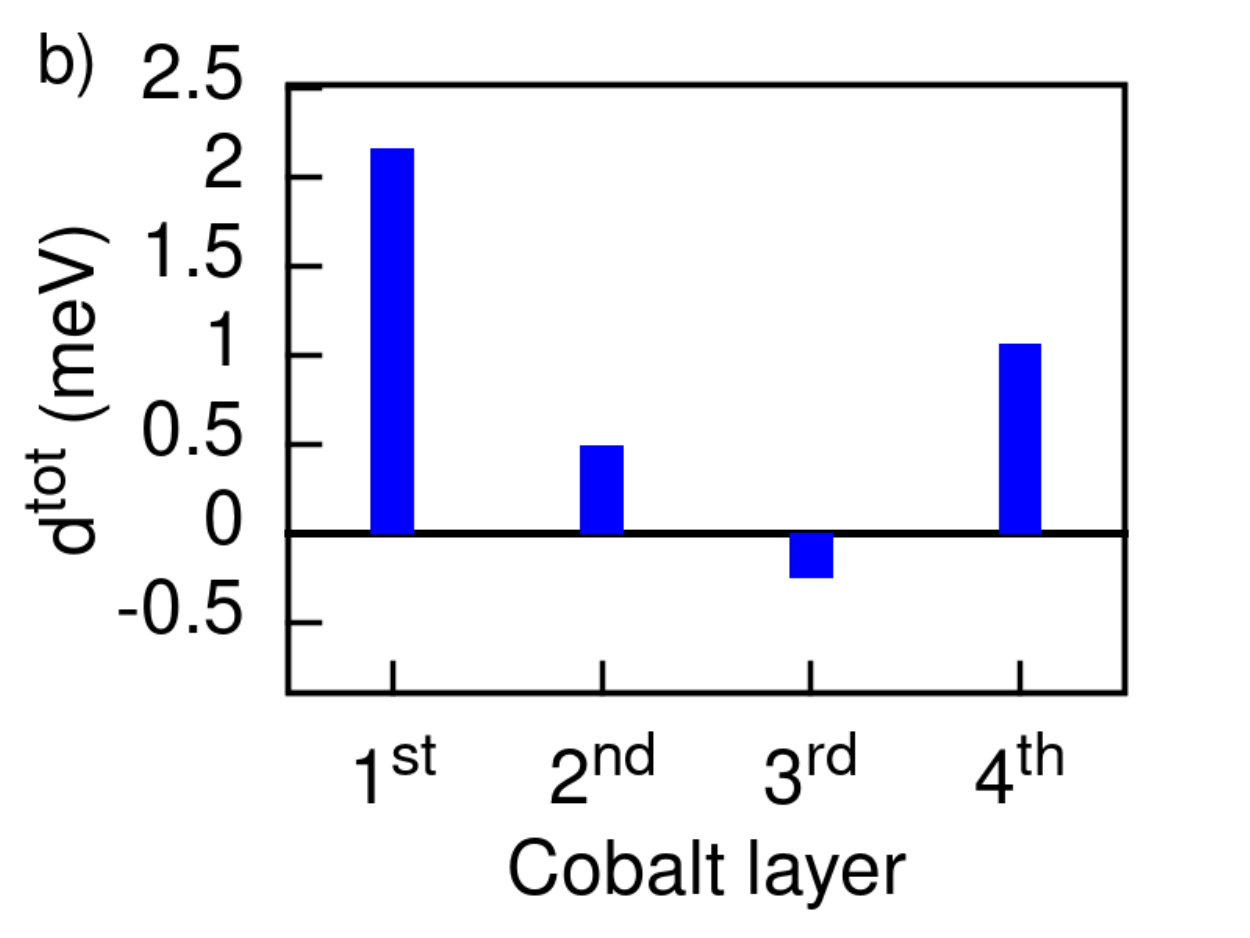}
\includegraphics[width=0.5\columnwidth,height=5cm]{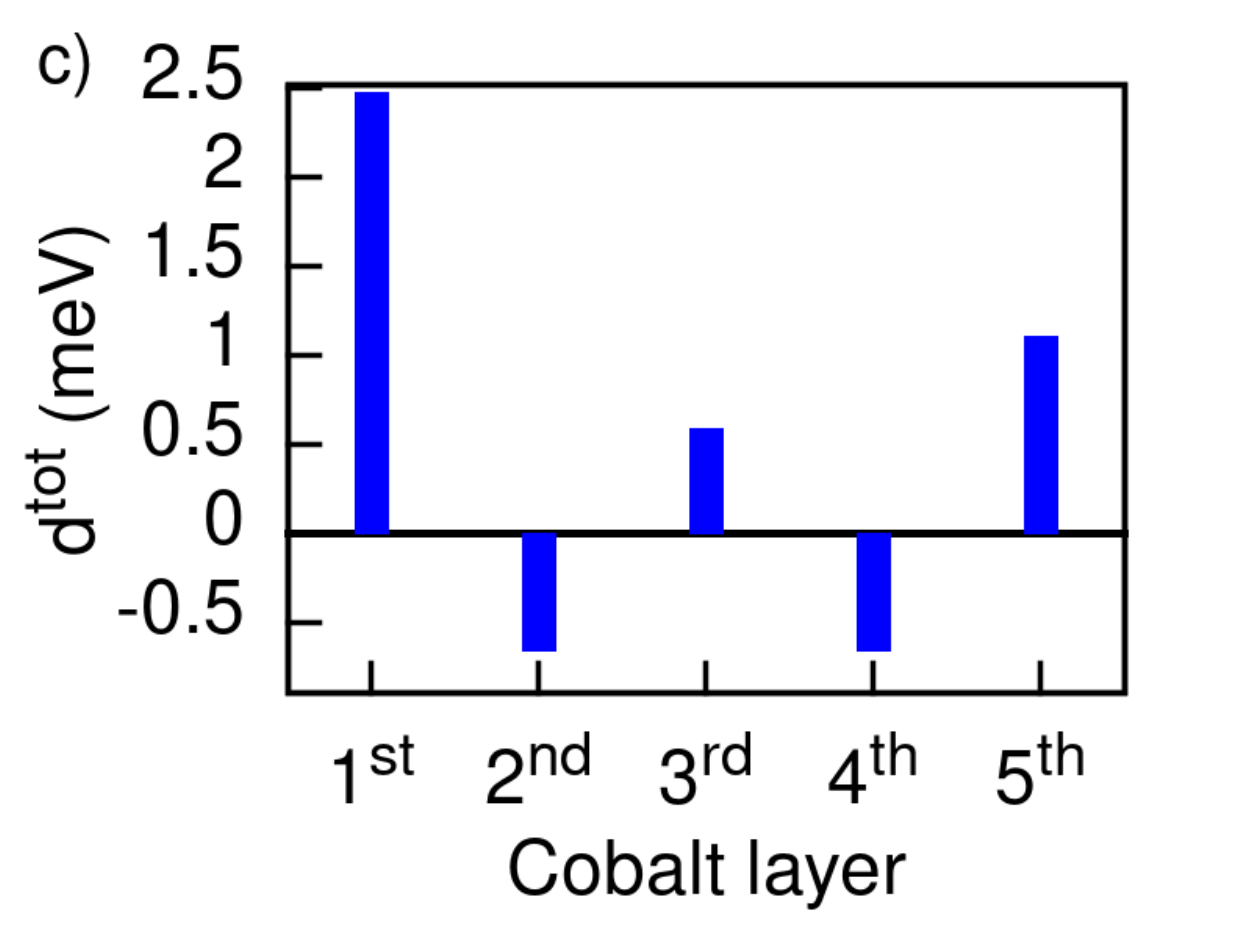}\includegraphics[width=0.5\columnwidth,height=5cm]{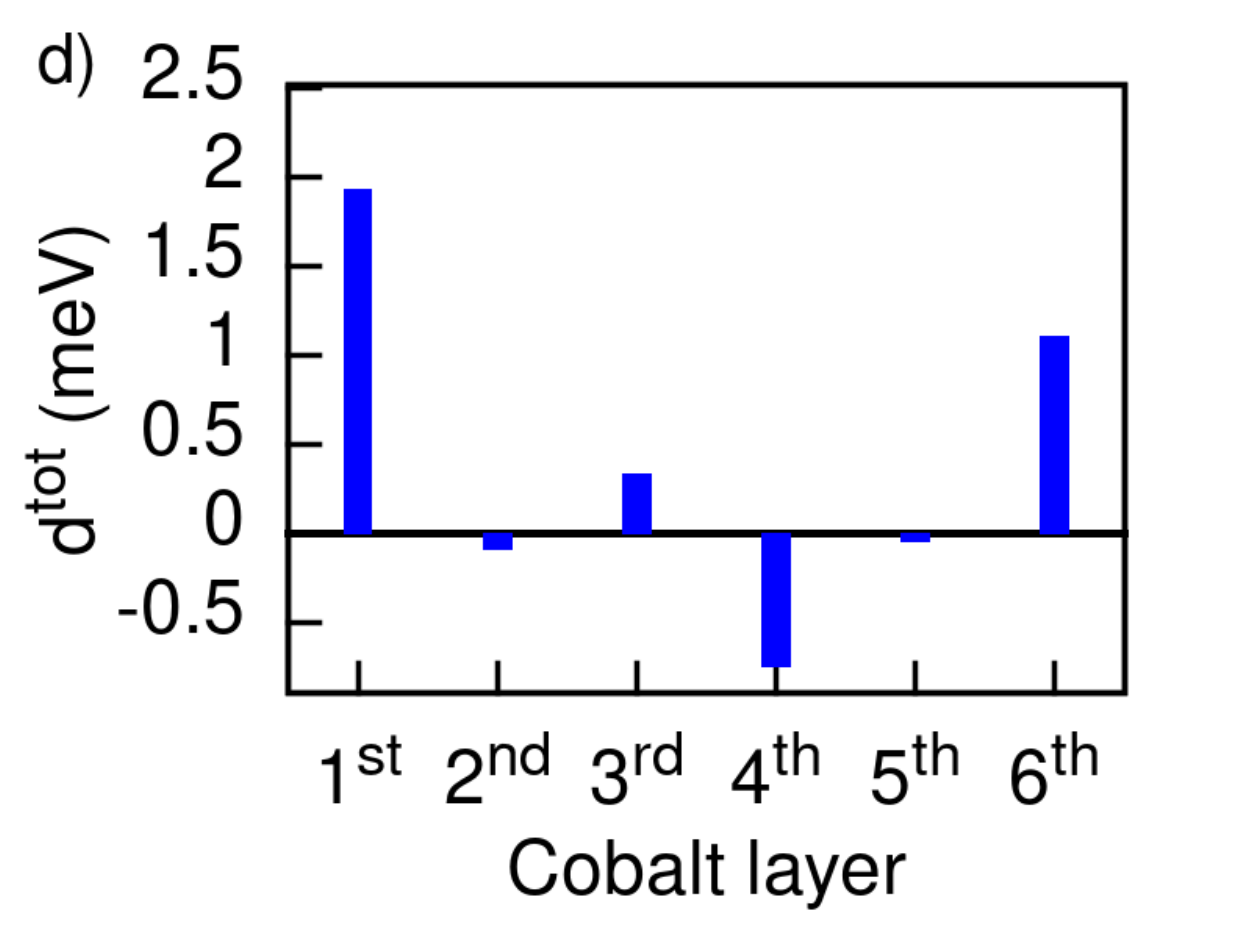}
\caption{Layer resolved DMI strength of the system with (a) for Pt$_5$/Co$_3$/Re$_5$, (b) Pt$_5$/Co$_4$/Re$_5$, (c) Pt$_5$/Co$_5$/Re$_5$ and (d) Pt$_5$/Co$_6$/Re$_5$. The first layer on the left side denotes the contribution coming from the Pt/Co interface while that at the right is due to Co/Re.}\label{resolved}
\end{figure}

The layer-resolved DMI strength on each cobalt layer was also calculated by introducing chiral magnetic moments for each layer separately. We calculate the layer-resolved DMI strength for Pt$_5$/Co$_n$/Re$_5$ with n=3, 4, 5 and 6 as can be seen from Fig. \ref{resolved}(a,b,c,d), respectively. For all cases, the calculated d$^k$ shows that two essential contributions are located at the Co interfaces with a positive sign. The other inner layers give a respectively smaller and oscillating {d$^k$} value which do not participate systematically to the total DMI. The oscillating nature of the DMI rises from the metallicity.\cite{PhysRevB.102.224427} The summation over $d^k$ should be close to the total value of the DMI, where the minor difference comes from the fact that a small DMI contribution comes from the Pt atoms. The value at the Pt side is around 2-2.5 meV, which is relatively high in comparison to the case of Pt/Co/W\cite{jena2021interfacial} system yielding half of the strength to around 1 meV. On the other hand, on the Re side, a value of around 1 meV was always achieved. The DMI vectors at the Pt and Re interfaces showing opposite chiralities are in agreement with experiments where the values at Pt/Co and Co/Re add up in the Pt/Co/Re systems producing a huge total DMI. It is also worth mentioning that the micromagnetic DMI depends on the magnetic layers’ thickness because DMI appears mainly at the interface as an effect of the hybridization between magnetic moments in the 3d cobalt layers and the strong SOC in 5d states of Pt and Re. If one has to compare the trilayer systems Pt/Co/Re with Pt/Co/W\cite{jena2021interfacial}, it is noticeable that the Pt/Co DMI contribution at the latter system is less in value. This behavior is similar to the case discussed in another work\cite{jia2018first}, where different stacks like Pt/Co/Rh and Pt/Co/Pd produced two distinct DMI values at the Pt/Co interface. Different transition metals with different SOC strengths can affect the overall DMI significantly.

\begin{figure}[t!]
\centering
\includegraphics[width=\columnwidth,height=5.9cm]{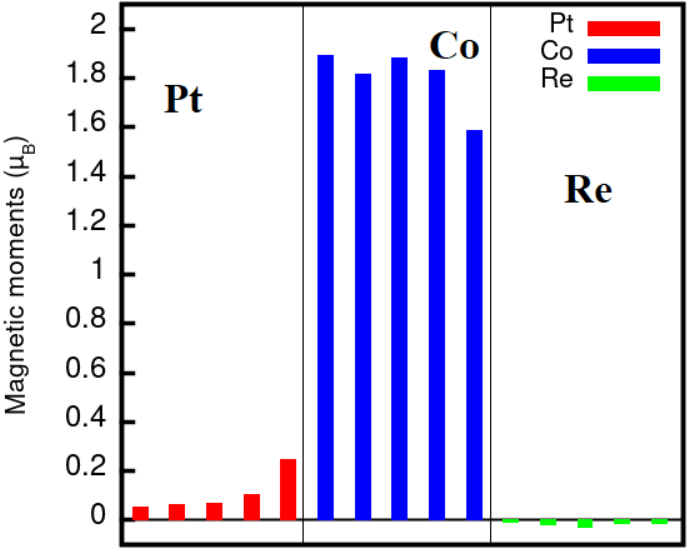}
\caption{Magnetization profile of the system Pt$_5$/Co$_5$/Re$_5$ with perfect interfaces. The system consists of 5 layers of each element in the stack. The interfacial magnetization effect weakly depends on the thicknesses of the Re, Co and Pt.}\label{perfect}
\end{figure}
The magnetization profile of the intrinsic spin of Co atoms and the induced magnetic moments at the 3d/5d interfaces were studied and displayed at equilibrium interlayer distances after performing the full relaxation scheme. This calculation was done for the ground state without SOC. The trend that is shown in Fig. \ref{perfect} was studied after that by varying the number of cobalt ALs. The figures show a positively induced spin polarization on the platinum atoms at the Pt/Co interface with a value of about 0.25 $\mu_B$. On the other hand, there is almost no visible polarization at the rhenium substrate at the Co/Re interface. The magnetic moments of cobalt dropped by 10\% of its original value, while that of rhenium shows very low polarizability which could be due to the weak hybridization of the 3d-states of cobalt with the low-lying states in rhenium, which is similar to the case for Co/Au\cite{belabbes2016hund} or Fe/Re\cite{autieri2016recipe}. By adding more cobalt ALs, the same behavior was seen even on relatively high thicknesses which confirms that rhenium is a weak polarizable transition metal.


To further investigate the influence of interfacial roughness on the DMI of the Pt/Co/Re structures, we generated a set of thin films with varying interfacial crystallinity by modifying the atomic arrangement of interfacial atoms. Two different types of mixed interface layers are presented in Fig. \ref{Structure}(b,c,d,e) denoted as C-mixing and Z-mixing which shows a different structural assortment of the atoms. Both the Pt/Co interface and the Co/Re interface were subject to this type of mixing, in order to study the influence of the different transition metals.

\begin{figure}[b!]
\centering
\includegraphics[width=\columnwidth,height=6.9cm]{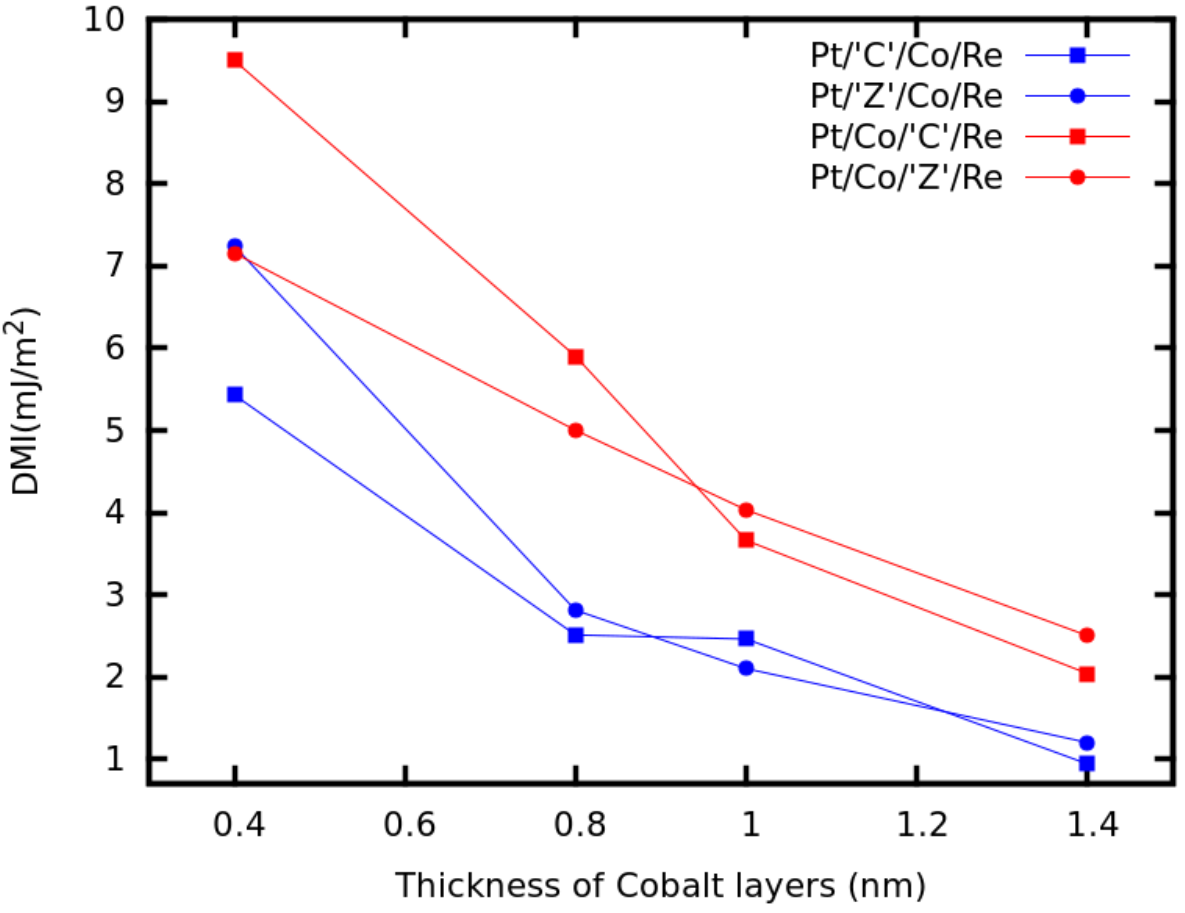}
\caption{Micromagnetic DMI results as a function of cobalt thickness for the two types of intermixing at both interfaces. The effective DMI here was calculated by dividing the total DMI by only the complete Co layers disregarding the mixed ones.}\label{DMI_Intermixing}
\end{figure}

\begin{figure*}[t!]
\centering
\includegraphics[width=\columnwidth,height=5.9cm]
{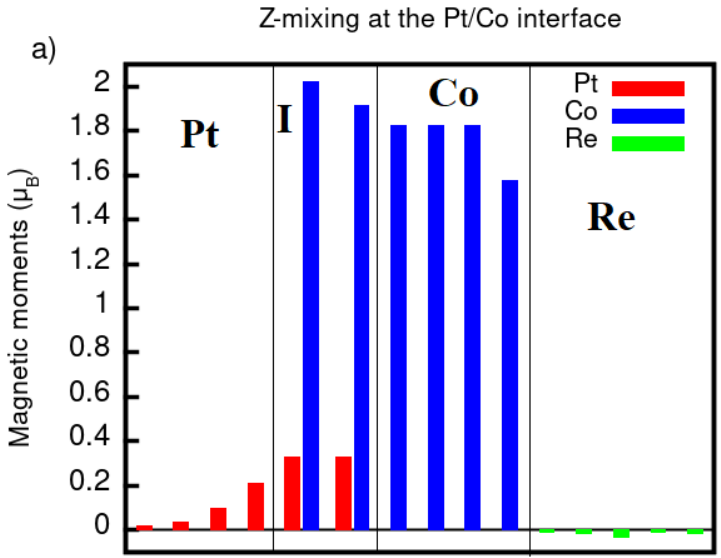}
\includegraphics[width=\columnwidth,height=5.9cm]
{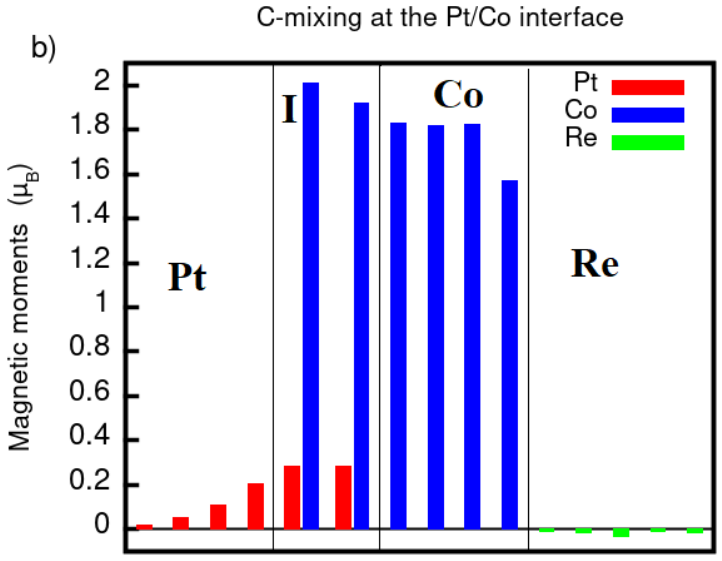}
\includegraphics[width=\columnwidth,height=5.9cm]
{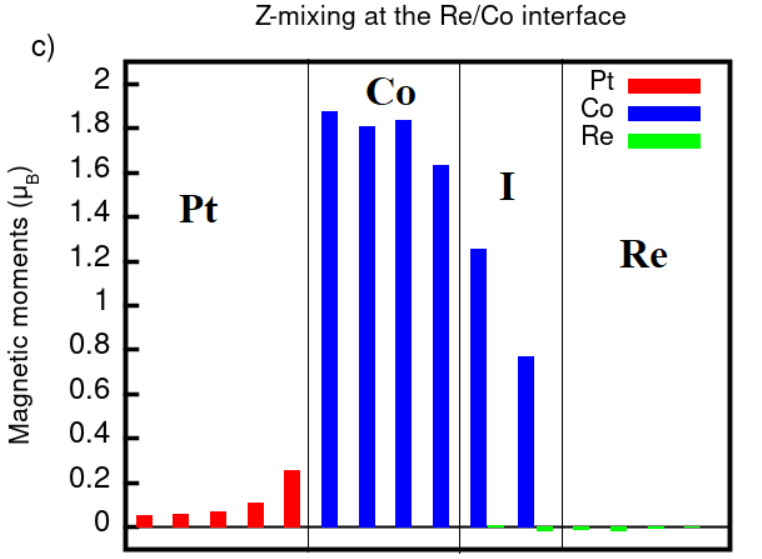}
\includegraphics[width=\columnwidth,height=5.9cm]
{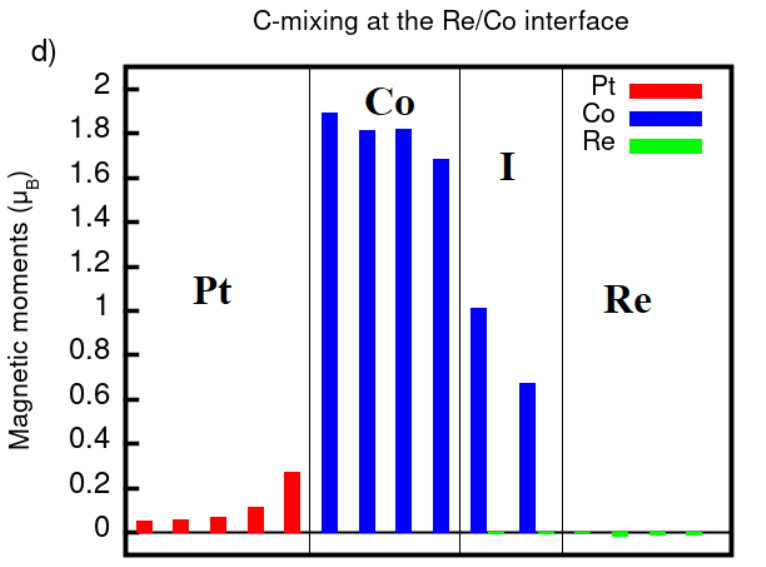}
\caption{Magnetization profile of the Pt$_5$/Co$_5$/Re$_5$ system with mixed interfaces: (a) Z- and (b) C-type mixing at the Pt/Co interface, (c) Z- and (d) C-type mixing at the Co/Re interface. We label the regions divided by vertical lines as Pt, Co, Re and I, where I indicates the intermixing region and Pt, Co and Re indicate the regions with ALs of a single element. After the mixing, these systems are such that the number of complete cobalt layers is 4 in addition to two mixed layers at the interface with half cobalt atoms.} \label{moments_mixed}
\end{figure*}

The method of the DMI calculations and the structural setup followed the same technique as in the case of perfect interfaces, where the spin spirals were included only on the cobalt atoms and on the other complete cobalt layers.  We begin our analysis by examining the situation in which the mixing is specifically implemented at the Pt/Co interface. By applying interfacial Co-Pt intermixing in either scenario, we found a decrease in the DMI which aligns with experimental findings suggesting that enhanced crystallinity leads to a larger DMI. With these kinds of intermixing, we do not have Co atoms first-neighbor in the intermixing region. It is expected to produce a sensitive reduction of the total DMI since the first-neighbor DMI is supposed to give the largest contribution to the total DMI. \\
The results of the micromagnetic DMI for the intermixing case are shown in Fig. \ref{DMI_Intermixing}. The effective DMI here was calculated by dividing the total DMI by only the complete Co layers disregarding the mixed ones. To initiate the analysis, the first studied system, which includes 1 cobalt layer (in the ideal case) and 2 cobalt layers (in the disordered case), displays a minor shift in the DMI value. Nevertheless, it's crucial to recognize that these cases are peculiar and cannot be classified as representative of the general case.
Moving to higher thicknesses, the value obtained for the system with two mixed layers of Co and Pt along with 3 complete layers, is 2.51 mJ/m$^2$ for the case of C-mixing and 2.81 mJ/m$^2$ in the case of Z-mixing. Moreover, the value for the system where we introduced two mixed layers of Co and Pt along with 4 complete layers, is 2.46 mJ/m$^2$ for the case of C-mixing and 2.10 mJ/m$^2$ for the case of Z-mixing. Finally, the case for two mixed layers of Co and Pt along with 6 complete layers of Co yielded  0.94 mJ/m$^2$ for the case of C-mixing and 1.20 mJ/m$^2$  for the case of Z-mixing. Therefore, the DMI was reduced by the disorder in all cases. Since the DMI between the second neighbors in the plane and between the first-neighbors out-of-plane is neglected, the cobalt atoms in the mixed layers do not contribute to the total DMI. Consequently, the DMI will depend on the last complete layer of Co-atoms. In the case of the mixed interface, the Co-atoms confront only 50\% of the Pt atoms, therefore, the interfacial DMI of the interfacial Co-layer is approximately reduced by 50\%.  However, we stress the role of the platinum atoms present in the disordered layers where fewer atoms participate in lowering the SOC of the system with respect to the ideal case. Therefore, giving an overall lower DMI value than that in the case of perfect interfaces. The experimental values of the micromagnetic DMI are expected to be between the values of the perfect interfaces and the intermixed ones. Coming to the case related to the Co/Re interface, the values of the DMI after intermixing were slightly smaller than the case of perfect interfaces as can be shown from Fig. \ref{DMI_Intermixing}. Since the Co/Re intermixing weakly affects the DMI, the DMI values for the Co/Re intermixing are larger than that for the Pt/Co intermixing.

After studying the DMI, we discuss the variation of the magnetic moments in the case of intermixing, especially at the interface where the coordination of cobalt atoms completely changes. We present the bar graphs in Fig. \ref{moments_mixed} in order to illustrate the effect. The figures present the system with 2 mixed interfaces in C and Z-type mixing, along with 3 complete cobalt layers and 5 layers for each of Pt and Re. 
Mixing at the Pt/Co interface increases the moments at Co atoms in both types of intermixing as shown in Fig. \ref{moments_mixed} (a,b) in the alloyed zone and also enhances the moment at the next Pt layer\cite{flores2022effect}.  The observed increase in the cobalt atoms aligns with the narrowing of the d bands, leading to an enhancement of magnetism caused by the progressive reduction of like-nearest neighbors\cite{zimmermann2018dzyaloshinskii}. When examining the Co/Re interface Fig. \ref{moments_mixed} (c,d), the magnetic moment of the Co interfacial layer experiences reduction which was also seen in the case of the ideal interface. Meanwhile, the magnetic moments generated within the Re layers display an inversion in its sign, being indicative of a possible antiferromagnetic coupling. However, its magnitude remains negligible, consistent with previously documented values\cite{simon2018magnetism}. Despite the changes observed in the magnetic moments of Co and Pt due to disorder, Re atoms exhibit no noticeable increase in polarizability, maintaining their state as in the ideal case.\\


Using first-principle calculations, we have proved that the chiral metallic multilayer Pt/Co/Re possesses a huge DMI interaction of up to micromagnetic DMI of 9.25 mJ/m$^2$ (or DMI strength of 4.19 meV). Re and Pt produce opposite chirality, therefore, the DMIs at the Co/Re and Pt/Co interfaces sum up to produce this huge effect.
The Re enhances the DMI at the Pt/Co interfaces. We have also found out that the induced magnetization is large and positive for the Pt up to 0.244 for the atoms interfaced with Co, while it is very small and negative for the Re atoms.  
We proved that Pt/Co/Re has better properties than the prototype system Pt/Co/W.
Upon study of the effect of intermixing on both interfaces, the intermixing on the Pt/Co interface introduces a suppression of the interfacial DMI, while, that on Co/Re shows very little or no effect on the system. Since the biggest contribution to the DMI comes from the Pt side of the stack, the intermixing at the Pt/Co interface produces a larger suppression of the total DMI.
Once the grown conditions are optimized to minimize the Pt/Co intermixing, the trilayer Pt/Co/Re could be used for skyrmions transport devices paving the way to new technological applications.\\

The authors were supported by the Polish National Science Centre under the project No. 2020/37/B/ST5/02299.
The work is supported by the Foundation for Polish Science through the International Research Agendas program co-financed by the European Union within the Smart Growth Operational Programme (Grant No. MAB/2017/1). 
We acknowledge the access to the computing facilities of the Interdisciplinary Center of Modeling at the University of Warsaw, Grant g91-1418, g91-1419 and g91-1426 for the availability of high-performance computing resources and support.
We acknowledge the CINECA award under the ISCRA initiative  IsC99 "SILENTS”, IsC105 "SILENTSG" and IsB26 "SHINY" grants for the availability of high-performance computing resources and support. We acknowledge the access to the computing facilities of the Poznan Supercomputing and Networking Center Grant No. 609.\\

\bibliography{DMI}
\end{document}